\documentclass[showpacs,twocolumn,pre]{revtex4}

\usepackage[dvips]{epsfig}

\newcommand{\clP}{{\cal P}}

\newcommand{\hL}{\hat{L}}

\newcommand{\PBr}{\{H,\dots\}_{PB}}

\newcommand{\prt}{\partial}

\newcommand{\rgl}{\rangle}
\newcommand{\lgl}{\langle}

\newcommand{\be}{\begin{equation}}
\newcommand{\ee}{\end{equation}}
\newcommand{\bea}{\begin{eqnarray}}
\newcommand{\eea}{\end{eqnarray}}

\begin{document}

\title{Dynamics of wave packets for the nonlinear
Schr\"odinger equation in random potential}

\author{Alexander Iomin}

\affiliation{Department of Physics, Technion, Haifa, 32000,
Israel}
\date{\today}

\begin{abstract}

The dynamics  of an initially localized Anderson mode is studied
in the framework of the nonlinear Schr\"odinger equation in the
presence of disorder. It is shown that the dynamics can be
described in the framework of the Liouville operator. An
analytical expression for a wave function of the initial time
dynamics is found by a perturbation approach. As follows from a
perturbative solution the initially localized wave function
remains localized. At asymptotically large times the dynamics can
be described qualitatively in the framework of a phenomenological
probabilistic approach by means of a probability distribution
function. It is shown that the probability distribution function
may be governed by the fractional Fokker-Planck equation and
corresponds to subdiffusion.

\end{abstract}


\maketitle

In this work the dynamics of an initially localized Anderson mode
is considered. It is described by the nonlinear Schr\"odinger
equation (NLSE) in the presence of disorder
\cite{bwFSW,bwDS,kivshar}. In the linear one-dimensional case,
eigenfunctions are localized \cite{Anderson,Lee}. This problem is
relevant to experiments in nonlinear optics, for example
disordered photonic lattices \cite{f1,lahini}, where Anderson
localization was found in the presence of nonlinear effects, as
well as to experiments on Bose-Einstein Condensates in disordered
optical lattices
\cite{BECE1,BECE3,akkermans,SanchAspect,BShapiro}. It was shown
that the presence of nonlinearity leads to an essential
complication of a mechanism of localization
\cite{bwFSW,bwDS,kivshar,if07}, and  the interplay between
disorder and nonlinear effects leads to new interesting physics
\cite{BECE3,akkermans,bishop1,CFK,dima,molina,Pikovsky,aubry,mackay,aub1}.
In particular, the problem of the spreading of wave packets and
transmission are not simply related \cite{doucot, pavloff}, in
contrast with the linear case. In spite of the extensive research,
many fundamental problems are still open. In particular,  the
spreading of an initially localized wave packet in nonlinear
finite chains was challenged in numerical studies with
realizations of subdiffusion \cite{dima,molina,Pikovsky} and
discrete breathers \cite{CFK,aubry,mackay,aub1,stavros}. The
absence of the wave packet diffusion was observed as well and this
behavior of initially localized wave packets was explained by
quasiperiodic solutions in the long time limit dynamics
\cite{stavros}.

The system under consideration is the NLSE
\be\label{am1}
 i\prt_t\psi=-\prt_x^2\psi+\beta|\psi|^2\psi+V\psi\, ,
\ee %
where $\beta$ is a nonlinearity parameter.  The variables are chosen in
dimensionless units and the Planck constant is $\hbar=1$. The
random potential $V=V(x), ~x\in(-\infty,+\infty)$ is such that for
the linear case $(\beta=0)$ the Anderson localization takes place,
and the system is described by the exponentially localized
Anderson modes (AM)s $\Psi_{\omega_k}\equiv\Psi_k(x)$, where
$\Psi_{\omega_k}(x)$ are real functions and the eigenspectrum
$\omega_k$ is discrete and dense \cite{LGP}. Therefore, the
problem in question is an evolution of an initially localized wave
function $\psi(t=0)=\Psi_{l_0}(x)$. Projecting Eq. (\ref{am1}) on
the basis of the Anderson modes
\be\label{am2}
 \psi(x,t)=\sum_{\omega_k}C_{\omega_k}(t)
\Psi_{\omega_k}(x)\equiv\sum_kC_k(t)\Psi_k(x)\, ,
 \ee %
we obtain a system of equations for coefficients of the expansion
$C_k$
\be\label{am3}
 i\prt_t{C}_k=\omega_kC_k+
\beta\sum_{k_1,k_2,k_3}A({\bf k}) C_{k_1}^*C_{k_2}C_{k_3}\, ,
 \ee %
where $A({\bf k}) $ is an overlapping integral
\bea\label{am4}
 A({\bf k})\equiv
A(k,k_1,k_2,k_3)   \nonumber \\
=\int\Psi_k(x)\Psi_{k_1}(x)\Psi_{k_2}(x)\Psi_{k_3}(x)dx\, .
 \eea %
The initial conditions for the system of Eqs. (\ref{am3}) are
\be\label{am5}
 C_k(t=0)=a_{\omega_k}\equiv a_k=\delta_{k,l_0}\, .
  \ee %
Equations (\ref{am3}) correspond to a system of interacting
nonlinear oscillators with the Hamiltonian
\be\label{am6}
 H=\sum_k\omega_kC_k^*C_k+\beta\sum_{\bf k}A({\bf
k})C_{k_1}^*C_{k_4}^*C_{k_2}C_{k_3}\, .
\ee %
Therefore, Eqs. (\ref{am3}) are produced by the Poisson brackets
$\PBr$ by means of the Liouville operator
\be\label{am7}
 \hL=\frac{1}{i}\PBr=\frac{1}{i}\left(\frac{\prt
H}{\prt{\bf C}_k^*}\cdot\frac{\prt}{\prt{\bf C}_k}-\frac{\prt
H}{\prt{\bf C}_k}\cdot\frac{\prt}{\prt{\bf C}_k^*}\right)(\dots)\,
.
 \ee %
Since $\hL H=0$ and $H(\{C_k,C_k^*\})=H(\{a_k,a_k^*\})$, we obtain
that the Liouville operator is an operator function of the initial
values:
\be\label{am7a}%
\hL=\frac{1}{i}\left[\frac{\prt H({\bf a}_k,{\bf a}_k^*)}{\prt{\bf
a}_k^*}\cdot\frac{\prt}{\prt{\bf a}_k} -\frac{\prt H({\bf
a}_k,{\bf a}_k^*)}{\prt{\bf a}_k}\cdot\frac{\prt}{\prt{\bf
a}_k^*}\right]
\ee%
and corresponds to the linear equation $\prt_t{\bf C}=\hL{\bf C}$.
Thus, the Liouville operator is the following combination:
\be\label{am7b}%
\hL=-i(\hL_0+\beta\hL_1)\, ,  
\ee%
where $\hL_0=\sum_k\omega_k\left(a_k\frac{\prt}{\prt
a_k}-{\rm c.c.}\right)$ and
\[    
\hL_1=\sum_{\bf k} A({\bf k})
\left[a_{k_1}^*a_{k_2}a_{k_3}\frac{\prt}{\prt a_{k_4}} -
{\rm c.c.}\right] \, .\]%
Here c.c. denotes a complex conjugation. Finally, we obtain that the
system of {\it nonlinear} ordinary differential equations
(\ref{am3}) is replaced by a system of {\it linear} partial
differential equations:
\be\label{am8} %
 \prt_t{C}_k(t)=\hL C_k(t)\, , ~~~k= 1,2,\dots ,\, .
 \ee %

A formal solution of Eq. (\ref{am10}) is the expansion
\be\label{am10_a} %
\bar{C}_k(t)=\sum_{n=0}^{\infty}\Big[\frac{t^n}{n!} \hL
^na_k\Big]_{a_k=\delta_{k,l_0}}\,
.\ee %
The nonzero contribution to the first power over $t$ of the
expansion (\ref{am10_a}) is due to the term
\be\label{am10_b} %
\hL_1^{(0)}=\sum_kA(l_0,l_0,l_0,k)|a_{l_0}|^2\left(a_{l_0}\frac{\prt}{\prt
a_k}-c.c.\right)\, , %
\ee %
while $(\hL_1-\hL_1^{(0)})a_k\equiv 0$ is due to the initial
conditions (\ref{am5}). Moreover, the contribution of
$\hL_1-\hL_1^{(0)}$ without $\hL_1^{(0)}$ is zero in all powers of
the expansion (\ref{am10_a}). For example, the $n$th power term
for $k\neq l_0$ is
\[
\left[\sum_{l\neq
l_0}A(l_0,l_0,l,l)|a_{l_0}|^2\prt_{\phi_l}\right]^{n}a_k
=i^nA^n(l_0,l_0,k,k)\delta_{k,l_0}\, .
\] %
It has a non zero contribution only in the power of the $n+1$
order after the action of the $\hL_1^{(0)}$ term \cite{add_june2}.
Therefore, keeping only the $\hL_1^{(0)}$ term in Eq. (\ref{am10})
means neglecting $O(\beta^2t^2)$ terms in the expansion
(\ref{am10_a}). This solution is valid up to a time scale
$t<t_{\beta}=1/\beta$.

To obtain a solution in the framework of this approximation, first
we eliminate the linear term $\hL_0$ from Eq. (\ref{am8}) by
substitution
\be\label{am9}
 \bar{C}_k(t)=\exp(-i\hL_0 t)C_k(t)\, .
  \ee %
After this substitution, Eq. (\ref{am8}) reads
\be\label{am10}
 \prt_t{\bar{C}}_k=
-i\beta\hL_1(t)\bar{C}_k,~~~\hL_1(t)=e^{-i\hL_0t}\hL_1e^{i\hL_0t}\, .
 \ee%
 Taking into account that
\[\exp[-i\hL_0t]=\exp\left[-\sum_k\omega_kt\frac{\prt}{\prt\phi_k}\right]\]
is the phase shift operator for the complex values
$a_k=|a_k|e^{i\phi_k}$, we obtain
 \be   \label{am11}%
 \hL_1(t)=\sum_{\bf k} A({\bf k})\left[       
\exp[i\Delta\omega t] a_{k_1}^*a_{k_2}a_{k_3}\frac{\prt}{\prt
a_{k_4}} -{\rm c.c.}\right]\, ,
 \ee %
where
$\Delta\omega=\omega_{k_1}+\omega_{k_4}-\omega_{k_2}-\omega_{k_3}$.

Solutions of Eq. (\ref{am10}) for $k\neq l_0$ are functions which
are zero at $t=0$. These are
\be\label{am12b} %
\bar{C}_{k}(t)=a_k+\frac{\beta
A_1|a_{l_0}|^2a_{l_0}}{\Delta\omega+\beta A_0|a_{l_0}|^2}
\cdot \left(e^{-i\beta A_0|a_{l_0}|^2t}-e^{i\Delta\omega
t}\right)\, .
 \ee %
Here $A_0=A(l_0,l_0,l_0,l_0)$ and $A_1\equiv
A_1(k)=A(l_0,l_0,l_0,k)\, , ~ k\neq l_0$, while $\Delta\omega$ now
is $\Delta\omega=\omega_k-\omega_{l_0}$. The complex conjugation
of Eq. (\ref{am12b}) is a solution as well. A solution for $k=l_0$
is a function of $\phi_{l_0}-\beta A_0|a_{l_0}|^2t$, which
corresponds to the initial conditions Eq. (\ref{am5}):
 \be\label{am12a}
\bar{C}_{l_0}(t)=a_{l_0}\exp(-i\beta A_0|a_{l_0}|^2t) \, .
\ee%
Using these analytical solutions for the coefficients
$\bar{C}_{k}(t)$ and Eq. (\ref{am9}), one obtains the solution of
the of NLSE (\ref{am1}) in the first order approximation over
$1/\beta$ as a sum
\bea\label{am13} %
&\psi(t)= a_{l_0}\exp(-i\omega_{l_1}t)\Psi_{l_0}(x)-
 4\beta |a_{l_0}|^2a_{l_0}\nonumber \\
 &\times\sum_k{}^{\prime}A_1(k)\frac{\sin
\left[\frac{\omega_k-\omega_{l_2}}{2}t\right]}{\omega_k-\omega_{l_2}}
 \sin\left[\frac{\omega_k+\omega_{l_2}}{2}t\right]
\Psi_k(x) \, , %
 \eea %
where $\omega_{l_1}=\omega_{l_0}+ \beta A_0|a|^2$ and
$\omega_{l_2}=\omega_{l_0}-\beta A_0|a|^2$, while prime means that
$k\neq l_0$. When $\beta\rightarrow 0$, we have at the
asymptotically large times $t_{\beta}\rightarrow\infty$ that
$\omega_{l_1}=\omega_{l_2}=\omega_{l_0}$, and the ${\rm sinc}$
function is
\[
\lim_{t\to\infty}
\frac{2\sin\left[\frac{\omega_k-\omega_{l_2}}{2}t\right]}
{\omega_k-\omega_{l_2}}
=2\pi\delta(\omega_k-\omega_{l_2})\, .\]%
The sum in Eq. (\ref{am13}) equals zero. Therefore, for $\beta=0$,
one obtains $\psi(t)=e^{-i\omega_{l_0}}\Psi_{l_0}(x)$ that
corresponds to a solution of the linear problem.

For nonzero values $\beta$ and $t<t_{\beta}$ the ${\rm sinc}$
function can be approximated by $t_{\beta}$ for
$\omega_k\approx\omega_{l_2}$. Then summation in Eq. (\ref{am13})
can be estimated by adding and subtracting the term with $k=l_0$.
Using the definition of the overlapping integrals $A_1(k)$ and
$\sum_k\Psi_k(x)\Psi_k(y)=\delta(y-x)$, one obtains an
approximation for Eq. (\ref{am13})
\be\label{am14} %
\psi(t)\sim \Psi_{l_0}(x)e^{-i\omega_{l_1}t}- 4\beta t
[\Psi_{l_0}^3(x)-A_0\Psi_{l_0}(x)]\sin(\omega_{l_2}t)
\, . %
\ee %
It means that at $t<t_{\beta}$ the wave function is localized and
its evolution corresponds to the periodic oscillations with the
frequencies $\omega_{l_1}$ and $\omega_{l_2}$. It is worth
mentioning that Eq. (\ref{am13}) is valid for both finite and
infinite systems (either discrete or continuous).

Consideration of the dynamics beyond $t>t_{\beta}$ relates to the
calculation of nonzero contributions of operators
$\Big[\hL_1-\hL_1^{(0)}\Big]^q$ and $\Big[\hL_1^{(0)}\Big]^p$,
acting on the initial conditions. This combinatorics leads to
essential difficulties for analytical treatment. To overcome this
obstacle the dynamics of the initially localized states can be
considered qualitatively in the framework of a phenomenological
probabilistic approach. To explain how the probabilistic approach
works, let us demonstrate it first for the localized solution of
Eq. (\ref{am14}). One obtains from the expansion (\ref{am10_a})
\be\label{am15} %
\sum_{n=1}^{\infty}\frac{(\beta t)^n}{n !}
\Big[A_0\frac{\prt}{\prt\phi_{l_0}}\Big]^{n-1}\Big[\hL_1^{(0)}\Big]\,
.
\ee %
The operator $\hL_1^{(0)}$ corresponds to the population of all
states $\Psi_k$ by transitions from the state $\Psi_{l_0}$. Since
all the states are localized at certain coordinates $X_k$, these
transitions correspond to ``jumps'' of a particle in the $x$
coordinate space from the position $X_{l_0}$ to positions $X_k$.
Therefore the operator $\hL_1^{(0)}$ corresponds to an instant
jump with the jump lengths distribution due to the  exponential
law in accordance with the overlapping integrals $A_1(X_k)$.
Another operator $A_0\frac{\prt}{\prt\phi_{l_0}}$ changes only the
phase of the complex amplitude $C_k$. Time duration of the action
of this operator is $t_{\beta}/n$, which is different for
different powers $n$. Now we introduce a probability distribution
function (pdf) $\clP(X,t)=|C_k(t)|^2$ to be a particle at position
$X$ at time $t$. Since the dynamics of the pdf $|C_k(t)|^2$ is
determined by the same Liouville operator as in Eq. (\ref{am7a}),
namely
\be\label{may1} %
 \prt_t |C_k(t)|^2=\hL |C_k(t)|^2\, , ~~~k= 1,2,\dots \, ,
 \ee %
we obtain that the pdf corresponds to the exponentially localized
solution \cite{add_may} $\clP(X,t)\sim
A_1^2(X)\sin^2\Big(\frac{\beta A_0t}{2}\Big)$ which is relevant to
Eq. (\ref{am14}) for $t<t_{\beta}$.

For $t\gg t_{\beta}$ Eq. (\ref{may1}) we present in the integral
form
\be\label{may2} %
 \clP(X,t)=\int_0^t\hL |C_k(t')|^2dt'\, .
 \ee  %
Since, in the new terminology, summation over indexes ${\bf k}$
corresponds to integration in space, the r.h.s. of Eq.
(\ref{may2}) can be rewritten in the form of the integral operator
\[\int_0^t\hL |C_k(t')|^2dt'\rightarrow \int_0^t
dt'\int_{-\infty}^{\infty} dX'\clP(X,t;X't')\clP(X',t')\, .\]
In this case all combinations of the overlapping integrals $A({\bf
k})$ with corresponding differentiating over $a_k$ play the role
of the kernel or transition probability $\clP(X,t;X't')$ of this
transformation. Therefore one has to consider a variety of
combinations of the operators
\[
\Big[A_0\frac{\prt}{\prt\phi_{l}}\Big]^{p_1}
\Big[\hL_1^{(0)}\Big]^{q_1}
\Big[\hL_1-\hL_1^{(0)}\Big]^{q_2}
\Big[A_0\frac{\prt}{\prt\phi_{k}}\Big]^{p_2}
\Big[\hL_1^{(0)}\Big]^{q_3}\dots
\, ,\] %
where $p_1+q_2+\dots+q_1+p_2+\dots=P+Q=n$. This corresponds to
different realizations of instant jumps and waiting times between
any two successive jumps. Note that
$\Big[A_0\frac{\prt}{\prt\phi_{l}}\Big]^{p_1}$ corresponds to
waiting time of duration $p_1t/P$. To proceed, we follow ideas of
so-called continuous time random walk \cite{mont,shlesinger}.
Therefore, the transition probability consists of the pdf of jump
lengths $f(X-X')$ and the pdf of waiting times $w(t-t')$. For
simplicity, we suppose that $\clP(X,t;X't')=f(X-X')w(t-t')$.

From the exponential decay of the overlapping integrals on the
large scale one obtains that all jump lengths have finite
expectation values and variances. Note, that
$\Big[\hL_1^{(0)}\Big]^{q_1} \Big[\hL_1-\hL_1^{(0)}\Big]^{q_2}$
corresponds to a ``jump'' which is a composition of random walks
(\textit{e.g.}, the simplest realization is presented in
\cite{add_june2}). For large $q_i$ the displacement
$\sum_{l=1}^{q_i}\Delta_l$ has Gaussian distribution due to the
central limit theorem. Here $\Delta_l=X_l-X_l^{\prime}$ are
transition lengths due to operator either $\hL_1^{(0)}$ or
$\hL_1-\hL_1^{(0)}$. Therefore, we can believe that these lengths
also are approximately distributed by the Gaussian law
$f(X-X')=\exp(- \Delta^2/2\sigma^2)/\sqrt{2\pi\sigma^2}$ and
$\sigma^2=\lgl\Delta^2\rgl$. It is worth stressing that the
overlapping integrals do not specify the pdf of waiting times.
Therefore, $w(\tau)$ can be defined from the average value of the
waiting times $\lgl\tau\rgl=\int_0^{\infty}\tau w(\tau)d\tau$.
This value also can be calculated from the following arguments.
For asymptotically large $t$, waiting times are $\tau=pt/P$, where
$p\in[1,P]$ and $P\in [1,n]$. Therefore, one obtains for the
average waiting time
\[
\lgl\tau\rgl=\lim_{n\to\infty}\frac{1}{n}
\sum_{P=1}^n\frac{1}{P}\sum_{p=1}^P\frac{pt}{P}=t\, .
 \]  %
This value diverges with $t\rightarrow\infty$, and it means that
there are infinitely many realizations of waiting times of the
order of $t$ \cite{add_june1}. To fulfill this condition, the
waiting times are distributed by power law $w(\tau)= \alpha
t_{\beta}/\tau^{1+\alpha}$ where $0<\alpha<1$, such that (see
\cite{add_june1})
\be\label{june2} %
\int_{t_{\beta}}^{\infty}\frac{\alpha
t_{\beta}^{\alpha}}{\tau^{1+\alpha}}d\tau =1~~~\mbox{and} ~~~
\int_{t_{\beta}}^{\infty}\frac{\alpha
t_{\beta}^{\alpha}}{\tau^{1+\alpha}}\tau d\tau =\infty\, .
 \ee  %
 It is reasonable to suppose that random
jumps and waiting times are independent and identically
distributed processes. Therefore this random qualitative
description of Markov operator $\hL$ in Eq. (\ref{am10_a}) and Eq.
(\ref{may2}) respectively, corresponds to the continuous time
random walk (see \textit{e.g}. Refs.
\cite{bouchaud,klafter,zaslavsky}) which is described by the
fractional Fokker-Planck equation
\be\label{ffpe} %
\prt_t\clP(X,t)-D_{\alpha}\prt_t^{1-\alpha}\prt_X^2\clP(X,t)=0\, ,
\ee %
where $D_{\alpha}=\sigma^2/t_{\beta}^{\alpha}$ is a generalized
diffusion coefficient and $\prt_t^{\nu}$ is a designation of the
Riemann-Liouville fractional derivative
\[\prt_t^{\nu}f(t)=\frac{1}{\Gamma(-\nu)}
\int_0^t\frac{f(\tau)d\tau}{(t-\tau)^{1+\nu}}\, .\] %
In this case only nonzero initial conditions can be taken into
consideration. Without restriction of the generality one can
consider $\clP(X,t=0)=\delta(X-X_{l_0})$. Equation (\ref{ffpe})
describes subdiffusion \cite{bouchaud,klafter,zaslavsky} since
$\alpha<1$. From Eq. (\ref{ffpe}) one obtains for the second
moment
\be\label{subdif} %
\lgl
X^2(t)\rgl=\int_{-\infty}^{\infty}X^2\clP(X,t)dX=D_{\alpha}t^{\alpha}\,
. \ee %
The transport exponent $\alpha$ cannot be specified here from the
developed probabilistic arguments. In the recent numerical studies
of the discrete NLSE \cite{Pikovsky,fks} the exponent $\alpha$ was
found in the range $0.3\div 0.4$.

In conclusion, the dynamics of the initially localized wave packet
$\psi(x,t)$ was studied in the framework of the NLSE in the
presence of a random potential. It is shown that the dynamics may
possibly be described in the framework the Liouville operator. The
interplay between disorder and nonlinearity $\beta$ leads to the
complicated dynamics of the initially localized state $\psi(x,
t=0)=\Psi_{\omega_0}(x)$. So, the influence of the nonlinearity on
the initial time dynamics is weak, and a perturbation theory in
$\beta$ can be developed. An analytical expression for a wave
function of the initial time dynamics is found by the perturbation
approach. As follows from a perturbative solution, at the initial
times $t<1/\beta$ the nonlinearity affects mainly the phase of the
wave function, while the shape of the wave packet corresponds to
the exponential localization due to the overlapping integrals
described by Eq. (\ref{am14}).

At asymptotically large times $t\gg 1/\beta$ the nonlinear effects
become important. To evaluate the influence of the nonlinearity on
the rate of spreading of the initial wave packet, one can consider
the large times asymptotic dynamics of the tails of the packet. In
this case, the dynamics can be described qualitatively in the
framework of a phenomenological probabilistic approach by means of
a probability distribution function $\clP(X,t)=|C_k(t)|^2$. The
last may be governed by the fractional Fokker-Planck equation
(\ref{ffpe}) which describes the asymptotic behavior of the tails
of the wave packet; its solution corresponds to subdiffusive
spread of the initially localized wave packet.

I am thankful to S. Fishman and S. Flach for helpful discussions
and critiques. The hospitality of the Max--Planck--Institute of
Physics of Complex Systems is kindly acknowledged. This research
was supported by the Israel Science Foundation.


\begin{thebibliography}{10}
\bibitem{add_june2} Another example explains the role of the operator
$\hL_1^{(0)}$. Let us define $\hL_1^{(1)}=\sum_{k_1,k_2\neq
l_0}A(l_0,l_0,k_1,k_2)|a_{l_0}|^2(a_{k_1}\frac{\prt}{\prt
a_{k_2}}-{\rm c.c.})$. Then we have $\hL_1^{(1)}a_k=\sum_{k_1\neq
l_0}|a_{l_0}|^2A(l_0,l_0,k_1,k)a_{k_1}=0$, and one obtains by
straightforward computation
$\Big[\hL_1^{(1)}\Big]^na_k=\sum_{\{k\}\neq
l_0}a_{k_1}|a_{l_0}|^{2n}$

$\times  A(l_0,l_0,k_1,k_2)A(l_0,l_0,k_2,k_3)\dots
A(l_0,l_0,k_n,k)=0\, .$

Finally, the nonzero contribution is due to the operator
$\hL_1^{(0)}$, namely
$\hL_1^{(0)}\Big[\hL_1^{(1)}\Big]^na_k=\sum_{\{k\}\neq
l_0}a_{l_0}|a_{l_0}|^{2(n+1)}$

$\times A(l_0,l_0,l_0,k_1)A(l_0,l_0,k_1,k_2) \dots
A(l_0,l_0,k_n,k)\, .$
\bibitem{add_may} We have from Eq. (\ref{am15})
$\int d\phi \frac{\prt}{\prt\phi}\sum_1^{\infty}\frac{(\beta
t)^n}{n!}[A_0\frac{\prt}{\prt\phi}]^{n-1}e^{i\phi}=
-ie^{i\phi}[e^{-i\beta A_0 t}-1]/A_0$.
\bibitem{add_june1} Introducing the maximal waiting time
$\tau_m(t)$ for the fixed $t$, one obtains
$\int_{t_{\beta}}^{\tau_m(t)}\tau
w(\tau)d\tau=t_{\beta}[\tau_m(t)]^{1-\alpha}\sim t$, and
$\tau_m(t)\rightarrow\infty$ in the limit $t\rightarrow\infty$. It
also follows that $\tau_m(t)\sim t^{\frac{1}{1-\alpha}}\gg t$.
\end{thebibliography}
\end{document}